\documentclass[5p,times,twocolumn,authoryear]{elsarticle}
\usepackage{graphicx}

\usepackage[version=4]{mhchem}
\usepackage[authoryear]{natbib}
\usepackage[colorlinks, citecolor=blue, linkcolor=blue, urlcolor=blue, breaklinks]{hyperref}
\usepackage{siunitx, xspace}
\usepackage{setspace}
\usepackage{csquotes}
\DeclareSIUnit[]\muB{\text{\ensuremath{\mu_{\textup{B}}}}}
\bibpunct{[}{]}{;}{n}{}{}

\newcommand{\uu}{1\hspace{-3pt}1}
\newcommand{\nag}{{\phantom{\dag}}}
\journal{~}
\usepackage{etoolbox}
\makeatletter
\patchcmd{\ps@pprintTitle}
  {Preprint submitted to}
  {~}
  {}{}
\makeatother

\begin{document}

\begin{frontmatter}

\title{A real-frequency solver for the Anderson impurity model based on
bath optimization and cluster perturbation theory}
\cortext[cor1]{Corresponding author}
\author[add1]{Manuel Zingl}
\address[add1]{Institute of Theoretical and Computational Physics,
Graz University of Technology, NAWI Graz, Petersgasse 16, 8010 Graz, Austria}
\author[add1]{Martin Nuss}
\author[add1]{Daniel Bauernfeind}
\author[add1]{Markus Aichhorn\corref{cor1}}
\ead{aichhorn@tugraz.at}

\date{\today}

\begin{abstract}
Recently solvers for the Anderson impurity model (AIM) working directly on the real-frequency axis have gained
much interest. A simple and yet frequently used impurity solver is exact diagonalization (ED), which is based on a
discretization of the AIM bath degrees of freedom. Usually, the bath parameters cannot be obtained directly on the
real-frequency axis, but have to be determined by a fit procedure on the Matsubara axis. In this work we present
an approach where the bath degrees of freedom are first
discretized directly on the real-frequency axis using a large number 
of bath sites ($\approx 50$). Then, the bath is optimized by unitary transformations such that it separates into
two parts that are weakly coupled. One part contains the impurity site and its interacting Green's functions can be
determined with ED. The other (larger) part is a non-interacting system containing all the remaining bath sites. Finally,
the Green's function of the full AIM is calculated via coupling these two parts with cluster perturbation theory.
\end{abstract}


\end{frontmatter}

\section{Introduction}
The single-orbital Anderson impurity model (AIM)~\cite{AIM} can be
represented exactly by an interacting site coupled to a bath of
infinitely many non-interacting sites. In approaches based on exact
diagonalization (ED), the number of sites in the interacting system is
restricted, and thus the bath needs to be
truncated~\cite{ED_caffarel,DMFT2, Capone_Lanczos}. This is a delicate
step, because no unique procedure exists. Different ways are used,
e.g., fits on the Matsubara axis or continuous fraction
expansions~\cite{DMFT2,EDtempbath, EDbathopt}. 

Various methods improving on ED have been presented in
recent years, e.g., the variational exact
diagonalization~\cite{VariationalED}, the distributional exact
diagonalization~\cite{DistributionalED} and methods based on a restriction
of the basis states~\cite{DMFT_QC, CI, DMFT_CI, Lu, Millis}.
Another way of going beyond ED is the use of cluster perturbation
theory (CPT)~\cite{CPT1,CPT2,CPTSene2002}, i.e. the more advanced
variational cluster approximation (VCA)~\cite{VCA,VCA1,VCA2}, as a
solver for the AIM~\cite{Nuss_VCA,CPT_DMFT}. 

\begin{figure}[t]
\centering
\includegraphics[width=0.7\columnwidth]{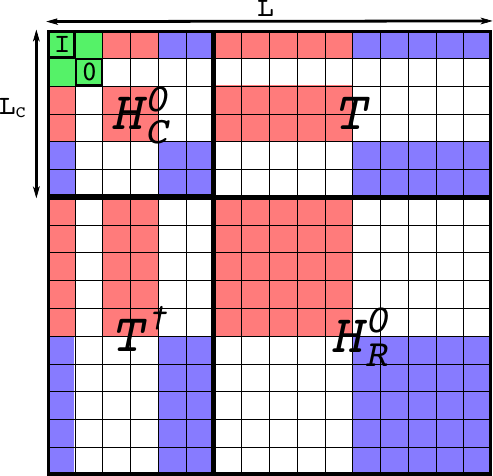}
\caption{Splitting of the Hamiltonian $H^0$ into an interacting cluster $H^0_C$ of $L_C=6$ sites, the remaining reservoir $H^0_R$ and the coupling of those two system $T$. In the specific example of a particle-hole symmetric system the unitary updates are performed in the space of negative energies (red) and equivalently in the space of positive energies (blue). No non-zero matrix elements are generated in the white blocks by the two-dimensional rotation matrices used in this work. The diagonal blocks from top to bottom represent: 1) the impurity, 2) the zero mode inside the cluster, 3) two negative modes inside the cluster, 4) two positive modes inside the cluster, 5) the remaining negative modes in the reservoir and 6) the remaining positive modes in the reservoir.}
\label{fig:sketch}
\end{figure}

From now on, we assume a single-orbital AIM coupled to a finite but
large bath of $L-1$ non-interacting sites. The basic idea of using CPT
as an impurity solver is to separate the $L$-site AIM into a cluster
of size $L_C$, which includes the impurity site and $L_C-1$ bath
sites, and a non-interacting system consisting of the remaining bath
sites. In general, the non-interacting Green's function is specified by
the Hamiltonian $H^0$, that is a matrix in orbital space of size $L
\times L$. For illustration purposes (see the sketch in
Fig.~\ref{fig:sketch}), we denote the upper left $L_C\times L_C$ block
in $H^0$ as the interacting cluster, subsequently $H^0_{C}$. The
remaining, lower $(L-L_C)\times (L-L_C)$ block describes the remainder
of the bath, subsequently $H^0_{R}$. Additionally, there are two off-diagonal 
blocks $T$ connecting $H^0_{C}$ and $H^0_{R}$. 
The onsite Hubbard interaction $H_U = U n_{I,\uparrow} n_{I,\downarrow}$,
where $I$ labels the impurity site, is now added to the cluster
Hamiltonian, $H_C=H_C^0+H_U$. There are no interactions in the bath
degrees of freedom, hence $H_R^0$ remains unchanged.

In CPT both Hamiltonians ($H_{C}$ and $H^0_{R}$) are solved exactly
for their single-particle Green's functions $G_C(\omega)$ and
$G_R(\omega)$. To obtain $G_C(\omega)$ we use the Lanczos procedure at zero temperature~\cite{Lanczos,ImpBandLanczos}. 
Note that $G_R(\omega) = G^0_R(\omega)$, as the
remainder of the bath is a non-interacting system. Subsequently, the
two systems are joined to yield the single-particle Green's function
of the full system $G(\omega)$ via the CPT relation~\cite{CPT2} 
\begin{align}
\label{eq:CPT}
 G^{-1}(\omega) &=
\left(\begin{array}{cc}
G_C(\omega) & 0  \\
0 & G_{R}(\omega)  \\
\end{array}\right)^{-1}-V\,\mbox{,}
\end{align}
where $V$ is a $L\times L$ coupling matrix consisting only of the $T$ blocks.
Eq.~\ref{eq:CPT} is exact in the case of a non-interacting system ($U=0$). In the
interacting case, the CPT relation is no longer exact, but a result of
perturbation theory in $V$. CPT approximates the self-energy of the
full system by the self-energy of the interacting cluster.  

In general, the non-interacting bath can always be transformed to a
tridiagonal representation via a Lanczos tridiagonalization, yielding
a chain representation of the AIM. This representation straight
forwardly allows to define the separation of the interacting cluster
and the remainder of the bath. However, the situation is not so clear in
other representations. Consider for example the case of a star
geometry, where all bath sites couple directly to the impurity site. 
Incorporating just a random set of these star sites into the
interacting cluster will lead to a poor discretization of the bath,
and hence a poor self-energy. 

Any unitary transformation on the non-interacting bath degrees of
freedom leaves the physics of the interacting AIM invariant. However,
such a transformation will influence the self-energy of the
interacting cluster significantly, since it changes the cluster
Hamiltonian $H_C$. Additionally, such transformations will also alter the
off-diagonal block $T$, rendering the resulting perturbation in some
cases larger than in others. There exist an infinite number of
representations which all describe the non-interacting bath exactly and
which are related via unitary transformations. However, the CPT method
itself suggests which baths might be the best: Those which
\enquote{minimize} the off-diagonal perturbative elements in $T$. The key
idea of this work is to use unitary transformations to find those
bath representations with minimal couplings between the cluster and the remainder of the bath.  

In the following, we outline a way to construct CPT-favorable bath
representations in Sec.~\ref{sec:method}, and present results for a $L=64$ AIM with a semi-circular particle-hole
symmetric bath in Sec.~\ref{sec:results}.

\section{Method}
\label{sec:method}

The general form of the non-interacting Hamiltonian for an $L$-site AIM is
\begin{align}
 H_0 &= \epsilon_{I} c^\dag_I c_I^\nag + \sum\limits_{i=1}^{L-1}
 \left(t_{iI} c^\dag_i c_I^\nag + {\rm h.c.}\right) +
 \sum\limits_{i, j = 1}^{L-1} t_{ij} c^\dag_i c_j^\nag\ ,
\end{align}
where the impurity is denoted by the index $I$ and the $L-1$ bath
sites by $i$ and $j$. We omit the spin indices. 
To obtain $H_0$ for an $L$-site system one can use a star
representation, where each bath site couples only to the impurity
site. Then, the parameters of $H_0$ can be determined by a
discretization of the non-interacting bath DOS into equally spaced
intervals. Each interval is represented by a delta peak, where the
energy positions of the delta peaks correspond to the on-site energies
and the hopping parameters are obtained from the spectral weights in the
intervals. Of course, the higher the number of bath sites the better the result
of this discretization.

Under a unitary transformation $R$, performed in the bath only, with
 $c_i^\nag =  \sum_{\alpha} R_{i\alpha} d_\alpha^\nag$ and $c_i^\dag = \sum_{\alpha} d_\alpha^\dag R^{*}_{\alpha i}$,
where $R R^\dag = \uu$, the transformed Hamiltonian reads
\begin{align}
 H'_0 &= \epsilon_{I} c^\dag_I c_I^\nag + \sum\limits_{\alpha=1}^{L-1}
 \left(h_{\alpha I} d_\alpha^\dag c_I^\nag + {\rm h.c.}\right) + 
 \sum\limits_{\alpha, \beta = 1}^{L-1} h_{\alpha\beta}
   d_\alpha^\dag d_\beta^\nag\ .
\end{align}
The parameters of the Hamiltonian transform like
 $h_{\alpha I} = \sum_{i} R^*_{\alpha i} t_{i I}$ and
 $h_{\alpha\beta} =\sum_{i,j} R^*_{\alpha i} t_{ij} R_{j\beta}$.
Such a transformation leaves the impurity state $I$ and consequently $\epsilon_{I}$ invariant.

We define an \enquote{energy} of a certain bath representation via the 2-norm of the off-diagonal blocks $T$
\begin{align}
 E = \frac{1}{N_T}\sum\limits_{i,j}|T_{ij}|^2\,\mbox{,}
\label{eq:bathe}
\end{align}
where the number of elements in $T$ is $N_T = L_C\cdot
(L-L_C)$. Transformations on the bath degrees of freedom included in
the interacting cluster do not influence the resulting
self-energy. The same is true for transformations performed only in the remainder of
the bath. This imposes a constraint on the energy $E$, namely, it has
to be invariant with respect to such transformations, which is indeed
fulfilled by the 2-norm. 

The aim is now to find an optimal bath representation
for CPT by minimizing the energy $E$. Since the configuration 
space of $T_{ij}$ is high dimensional, we use 
a Monte Carlo procedure. Initially, we perform
global updates in all dimensions with random
rotation matrices to obtain a randomized starting representation of
$H_0$. Then, we move through the space of possible $H_0$ by proposing
random local updates $R$. In general, any unitary update would be allowed, 
but here we restrict ourselves to two-dimensional rotation matrices for the 
local updates 
\begin{align*}
\setlength\arraycolsep{2pt}
 R = \left(\begin{array}{ccccccccc}
1 & 0 &  \ldots & 0 & 0 & \ldots & 0 & \ldots& 0\\
0 & 1 &  \ldots & 0 & 0 & \ldots & 0 & \ldots& 0\\
\ldots  & \ldots &  \ldots & \ldots & \ldots & \ldots & \ldots & \ldots& 0\\
0 & 0  &  \ldots & \cos\left(\phi\right)  & 0 & \ldots & -\sin\left(\phi\right)& \ldots& 0\\
0 & 0  &  \ldots & 0  & 1 & \ldots & 0& \ldots& 0\\
\ldots & \ldots &  \ldots & \ldots  & \ldots & \ldots & \ldots& \ldots& 0\\
0 & 0  & \ldots & \sin\left(\phi\right)  & 0 & \ldots & \cos\left(\phi\right)& \ldots& 0\\
\ldots & \ldots &  \ldots & \ldots & \ldots & \ldots & \ldots & \ldots& \ldots\\
0 & 0 &  \ldots & 0 & 0 & \ldots & 0 & \ldots & 1\\
\end{array}\right)
\end{align*}

A local update matrix $R(i,j,\phi)$ is drawn by choosing two random
integers $i,j \in [1,L-1]$ representing the plane of rotation and one
rotation angle $\phi \in [0,2\pi[$. A new
representation with energy $E'$ is accepted with probability $p =
\text{min}\left(1, e^{-\gamma(E'-E)}\right)$. We use simulated
annealing to obtain low-energy CPT bath representations by increasing
the parameter $\gamma$. 

Although bath rotations leave the particle-hole symmetry invariant on
the $L$-site $H_0$, they destroy it on the $L_C$-site cluster. Therefore, 
as shown in Fig.~\ref{fig:sketch}, we split the bath sites into an equal 
amount of positive (blue elements)
and negative energy (red elements) sites and one zero mode (green
0). Updates are performed simultaneously on the positive and negative
modes which leaves the whole bath, the bath in the cluster as well as
the remaining bath particle-hole invariant.
To avoid a Kramers-degenerate ground state, clusters with an even
number of sites $L_C$ are chosen. This implies that one bath site (the
zero mode) is exactly located at zero energy. Zero mode updates cannot be achieved by two-dimensional rotations without breaking the
particle-hole symmetry of the cluster, but would rather require
special unitary transformations involving at least three bath
sites. For the proof of concept presented here, we refrain from updating 
the zero mode, i.e. the green elements in Fig.~\ref{fig:sketch} do not change.
Hence, the zero mode coupling is determined by the initial
discretization of the system. Although this restricts the space of
trial bath representations, we leave the zero mode updates for future
works.  

Next to the energy $E$ (Eq.~\ref{eq:bathe}), which reflects the
magnitude of the perturbation, we evaluate the influence of the CPT
truncation by comparing the non-interacting single-particle impurity
Green's function of the full system $G^{I}_0(\omega)$ to the one
considering only the sites in the cluster $G^{I}_{0,C}(\omega)$. The
resulting quantity 
\begin{align}
\label{eq:deltachi}
\chi^2_{C}=\int d\omega \left|G^{I}_0(\omega+i\eta)-G^{I}_{0,C}(\omega+i\eta)\right|^2\,\mbox{,}
\end{align}
reflects the ability of the $L_C$ cluster sites to represent the bath
degrees of freedom. A-priori a positive correlation of $E$ and
$\chi^2_{C}$ is not ensured but expected. We emphazise that
$\chi^2_{C}$ is not used in the algorithm, but only serves as a
measurement for the quality of the bath optimization. A numerical broadening of $\eta =\SI{0.02}{\electronvolt}$ is used to evaluate Eq.~\ref{eq:deltachi}.

To asses the quality of the optimization scheme we also perform plain ED calculations for a truncated $L=10$ system. The Hamiltonian of this 10-sites system (with the Green's function $G^I_{0,ED}(i\omega_n)$) is obtained by fitting $G^I_0\left(i\omega\right)$ on the Matsubara axis with the cost function 
\begin{align}
\label{eq:deltachiED}
\chi^2_{ED}=\sum \limits_{n} W_n \left|G^I_0(i\omega_n)^{-1}-G^I_{0,ED}(i\omega_n)^{-1}\right|^2\,\mbox{.}
\end{align}
We employ the simplex search method by Lagarias {\em et al.}~\cite{Lagarias}
and impose particle-hole symmetry to reduce the number of fit
parameters. A Matsubara grid with 1024 points at a fictitious
temperature corresponding to $\beta =\SI{100}{\electronvolt^{-1}}$ is
used, but the ED solution itself is obtained at zero temperature. 
The cost function Eq.~\ref{eq:deltachiED} is a heuristic choice and can also
take various other forms, e.g. with a different definition of the
distance or a different weight $W_n$ (we set $W_n = 1/\omega_n$)~\cite{EDtempbath, EDbathopt} . 
This leads to an ad-hoc determination of the bath
parameters and introduces some ambiguity to the solution of the
AIM. For a given number of bath sites also the discretization on the real-frequency axis 
is not uniquely defined, e.g. we could use unequally spaced energy intervals. 
In contrast to ED, where usually only a small number of bath sites is fitted on the Matsubara axis via the
minimization of Eq.~\ref{eq:deltachiED}, the discretization on the real-frequency axis can be easily 
performed for a large number of sites. Indubitably, the ambiguities in determining the AIM parameters are less severe for larger system sizes.

\section{Results}
\label{sec:results}
Here, we discuss a single interacting impurity in a particle-hole symmetric semi-circular bath with a half-bandwidth of $\SI{1}{\electronvolt}$. For the bath optimization we use a discretized system with a total number of $L=64$ sites and an interacting cluster of $L_C=10$. The interacting cluster includes the impurity site, one zero mode, and four additional positive and negative bath sites each. 

\begin{figure}[t]
\centering
\includegraphics[width=0.93\columnwidth]{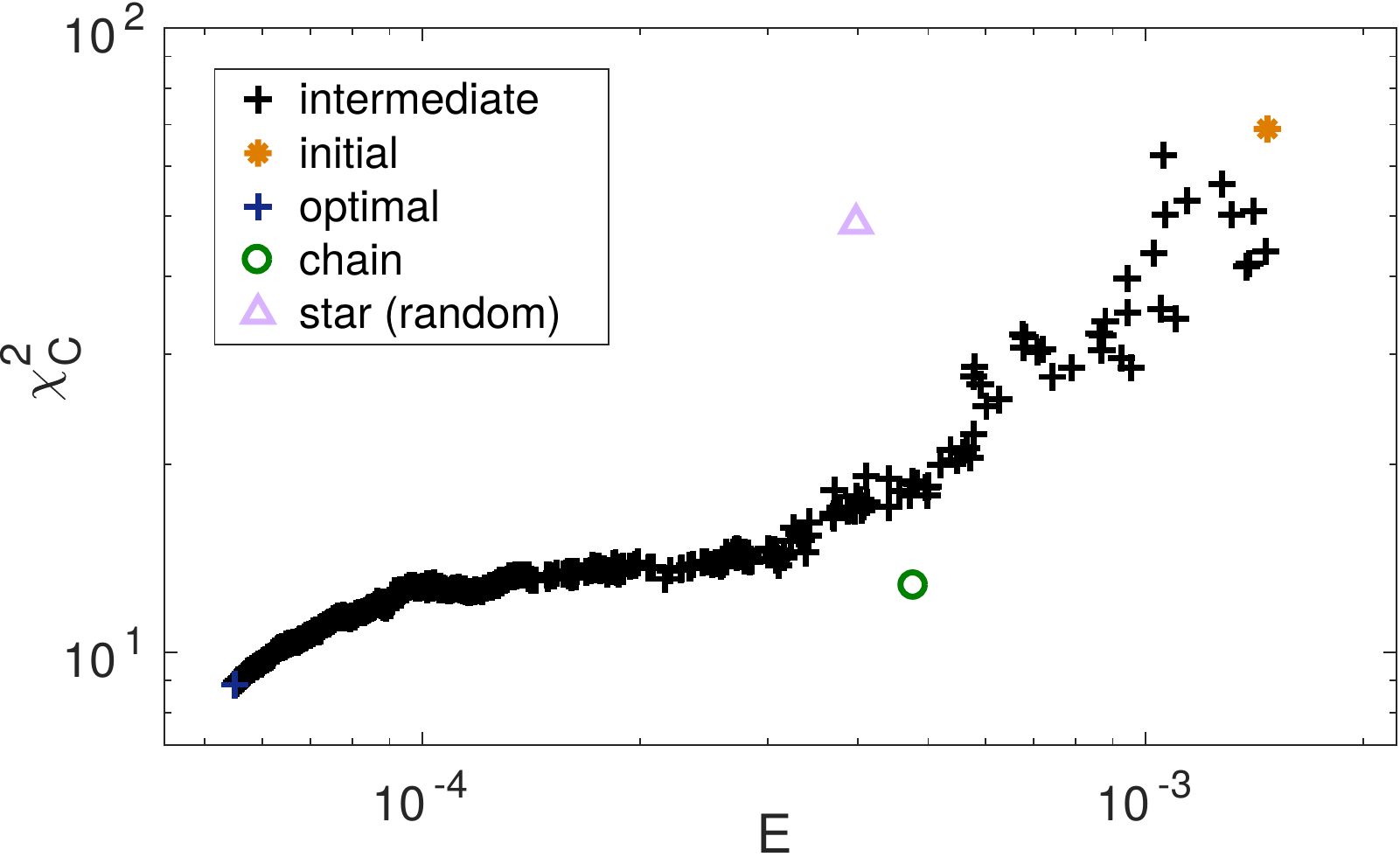}
\caption{Correlation of the cost function $\chi^2_{C}$ and the Monte Carlo energy $E$ at intermediate steps of the optimization procedure (black crosses). $\chi^2_{C}$ measures the quality of the representation of the non-interacting full system by the 10-site cluster. In addition, we show the random starting point (orange star), the optimized representation (blue cross), the respective values of a chain representation cut after 10 sites (green circle) and an arbitrary selected 10-site star (cyan triangle).}
\label{fig:deltaChi}
\end{figure}

Fig.~\ref{fig:deltaChi} shows that smaller perturbative elements $E$ correlate positively with smaller $\chi^2_{C}$, indicating a better representation of the non-interacting bath by the sites contained in the interacting cluster. We compare intermediate bath representations (black crosses) to the random initial representation (orange star), a star representation by choosing ten sites at random to enter the interacting cluster (cyan triangle), and the chain representation cut after 10 sites (green circle). Note that the chain representation hosts only one perturbative matrix element, which is however large. In all cases one finds a higher $\chi^2_{C}$ with respect to the final result of the optimization (blue cross).  Of course, due to the optimization the number of elements in $T$ grows, but their magnitude becomes tiny compared to the chain or the star representation.

\begin{figure}[t]
\centering
\includegraphics[width=0.93\columnwidth]{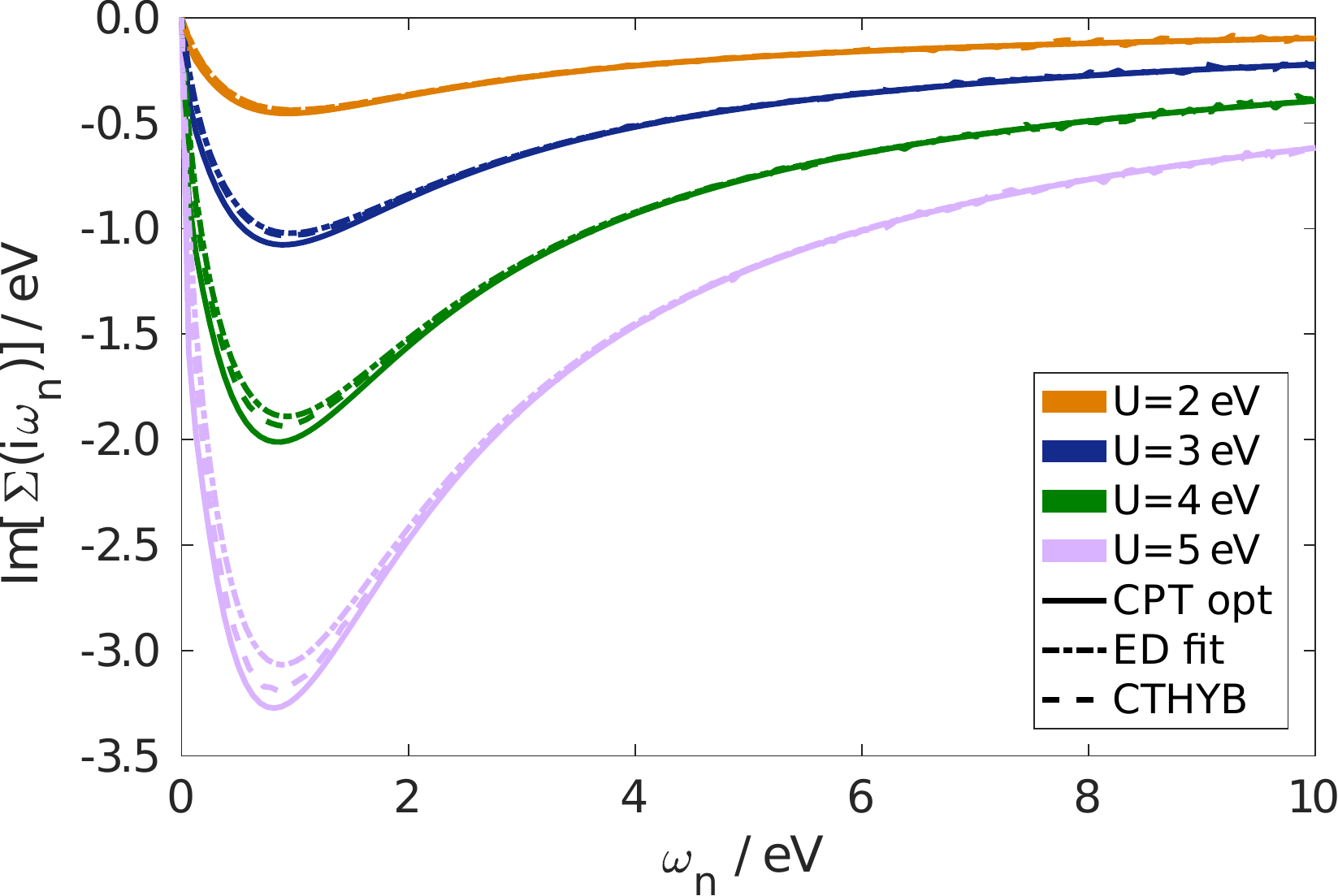}
\caption{Matsubara self-energy of the optimized 10-site cluster (full lines) compared to the ED self-energy of a fitted 10-site system (dash-dotted lines) and a CTHYB result using the continuous bath DOS (dashed lines). The self-energy is shown for interaction values of $U= 2, 3, 4 \text{ and } \SI{5}{\electronvolt}$.}
\label{fig:Sigma}
\end{figure}

In Fig.~\ref{fig:Sigma} the self-energy $\Sigma(i\omega_n)$ of the optimized system on the Matsubara axis is shown for different values of $U$. We compare it to results obtained at an inverse temperature of $\beta=\SI{100}{\electronvolt^{-1}}$ with the TRIQS/CTHYB solver~\cite{TRIQS/CTHYB}, which is based
on continuous-time quantum Monte Carlo in the hybridization expansion (CTHYB)~\cite{CTQMC, werner2}. Additionally, we show the self-energies calculated with plain ED for a $L=10$ system.
As the CPT self-energy of the full AIM is given by the self-energy of the 10-site 
cluster, we can only expect it to be compatible with CTHYB on a similar level as the ED self-energies are. For $U<=\SI{4}{\electronvolt}$ the ED 
self-energies are even slightly closer to the CTHYB result (see Fig.~\ref{fig:Sigma}).
Nevertheless, the important message conveyed by Figs.~\ref{fig:deltaChi} 
and~\ref{fig:Sigma} is that the minimization of the off-diagonal
block $T$ is a proper procedure to obtain a good representation of the 
full AIM system by the 10-site cluster.

\begin{figure}[th]
\centering
\includegraphics[width=0.95\columnwidth]{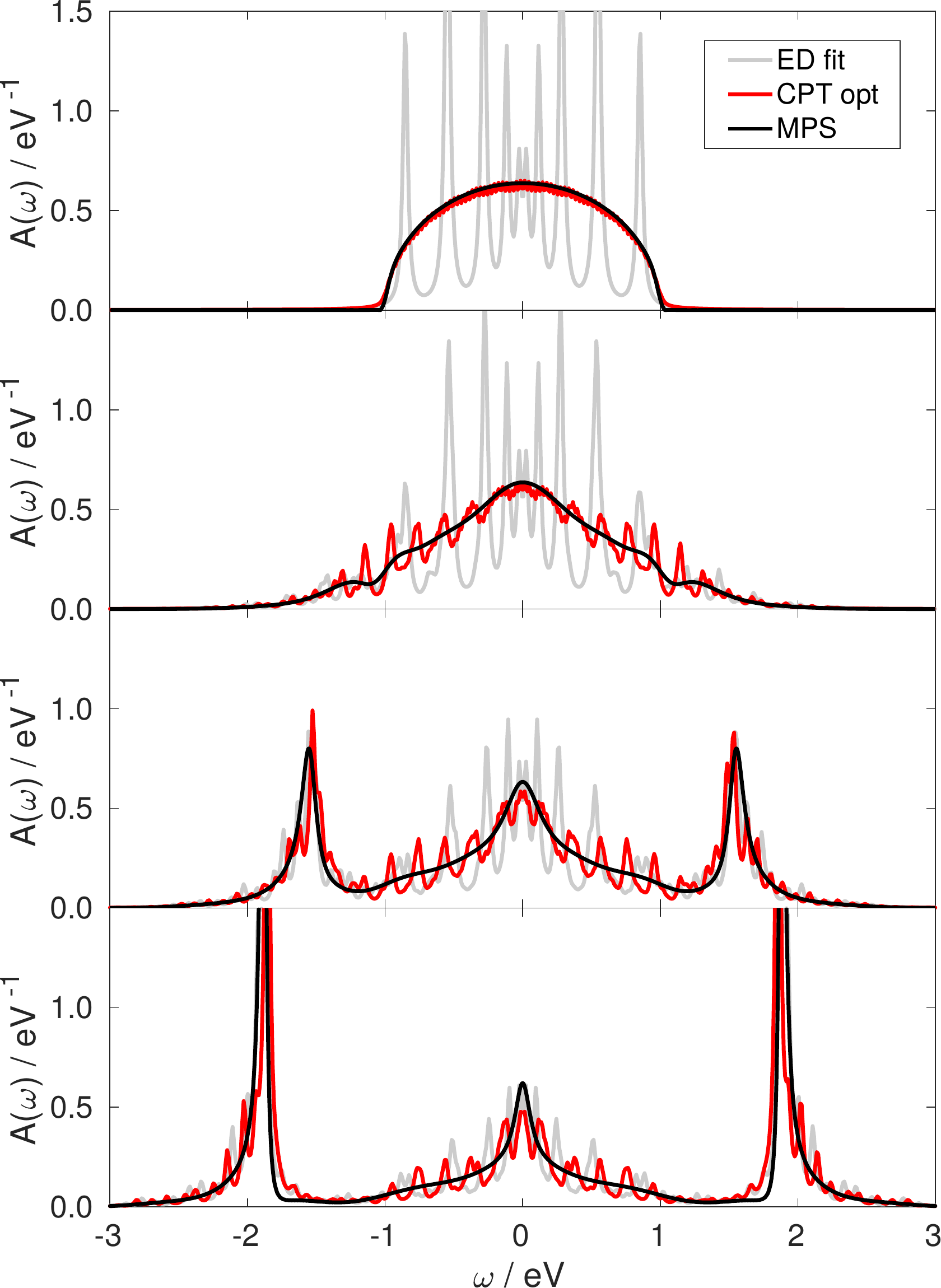}
\caption{Spectral function $A(\omega)$ of the AIM with a particle-hole symmetric semi-circular bath DOS. The CPT interacting spectral functions $A(\omega)$ of the optimized system of size $L=64$ and $L_C = 10$ (red) is compared to MPS results for the $L=64$ system in star representation (black) and to the ED result for a 10-site system (gray). A numerical broadening of $\eta = \SI{0.02}{\electronvolt}$ was used to calculate the ED and CPT spectral functions.}
\label{fig:G}
\end{figure}

The actual advantage of CPT is revealed on the level of the spectral function $A(\omega)$, which results from coupling the interacting cluster 
to the remainder of the non-interacting bath.
In Fig.~\ref{fig:G} we show the CPT spectral function (red) of the
optimized system ($L=64$) for interaction values $U=0, 1, 2,
\text{ and } \SI{3}{\electronvolt}$. We compare our results to spectral
functions obtained with the density matrix renormalization group (DMRG)~\cite{WhiteDMRG,SchollwoeckDMRG_MPS} and real-time evolution (as in 
Ref.~\cite{FTPS}) using matrix product states (MPS), colored in black. 
This approach is able to provide an excellent spectral resolution on the whole real-frequency axis~\cite{FTPS, WolfStar,MartinTEBD}. We use the star representation of the $L=64$ AIM for the MPS calculation with a truncated weight of $10^{-10}$. Additionally, we show the results of the 10-site ED system (gray).
In contrast to ED, which exhibits strong finite size effects, the CPT optimization scheme provides smoother spectral functions which are in much better agreement with the MPS results. This difference is particularly pronounced for the smaller 
$U$ values, which is a consequence of CPT becoming exact for $U\rightarrow 0$.
Up to the influence of the discretization, CPT
reproduces the non-interacting spectral function for $U=\SI{0}{\electronvolt}$ 
(see top grap in Fig.~\ref{fig:G}). Of course, for higher values of $U$ the energy resolution is limited by the size of the exactly solved system ($L_C$), which can be improved on by including more sites in the interacting cluster.

\section{Conclusion}
In this work we introduced a bath optimization scheme for the Anderson
impurity model. Using unitary transformations in the bath degrees of
freedom, we minimize the coupling between a small cluster
containing the interacting impurity site and the remaining sites of the
bath. These transformations leave the impurity DOS of the
non-interacting bath invariant. In general, the proposed scheme can be
useful for all CPT-based methods when parts of the entire system are
non-interacting, but it does in principle also provide a guideline to
construct finite-size representations of hybridization functions as
needed, e.g., in the framework of dynamical mean-field theory. For a
large enough number of bath sites, the initial AIM can be obtained
directly on the real-frequency axis, and thus a fit on the Matsubara
axis can be avoided. In this work we have presented a proof of
concept, but anticipate to explore the bath optimization scheme for
systems without particle-hole symmetry and multi-orbital impurity models.

\section*{Acknowledgments}
The authors acknowledge financial support from the Austrian Science Fund
(FWF) through SFB ViCoM F41 (P04 and P03), project P26220, and through 
the START program Y746, as well as from NAWI-Graz. The CTHYB results were calculated using the TRIQS library~\cite{TRIQS} and the TRIQS/CTHYB solver~\cite{TRIQS/CTHYB}. 
The MPS results were obtained using the ITensor library~\cite{ITensor}.
\section*{References}
\bibliographystyle{elsarticle-num}
\bibliography{literatur.bib}
\end{document}